\documentclass[aps,prd,twocolumn]{revtex4}
\pdfoutput=1
\usepackage{textcomp}
\usepackage{amssymb}
\usepackage{bold-extra}
\usepackage{enumerate}
\usepackage{graphicx}
\usepackage{mathrsfs}
\usepackage{multirow}
\usepackage{rotating}
\usepackage{subfig}
\usepackage{url}
\usepackage{wrapfig}
\usepackage{amsmath}
\usepackage{color}
\usepackage{verbatim} 

\def\nn{\nonumber}
\def\bea{\begin{eqnarray}}
\def\eea{\end{eqnarray}}
\def\ba{\begin{eqnarray}}
\def\ea{\end{eqnarray}}
\def\be{\begin{equation}}
\def\ee{\end{equation}}
\def\beq{\begin{equation}}
\def\eeq{\end{equation}}

\begin{document}

\title{Bottom-Quark Forward-Backward and Charge Asymmetries at Hadron Colliders}
\author{Christopher W. Murphy}
\email{christopher.murphy@sns.it}
\affiliation{Scuola Normale Superiore, Piazza dei Cavalieri 7, Pisa 56126 Italy}

\begin{abstract}
Predictions are made for the forward-backward and charge asymmetries in bottom-quark pair production at hadron colliders. Tree-level exchanges of electroweak (EW) gauge bosons dominate the Standard Model (SM) contribution to the asymmetry near the $Z$-pole. The mixed EW-QCD corrections are computed in an approximate way, and are found to be small in magnitude. These SM predictions are consistent with experimental results from CDF, D0, and LHCb. In particular, CDF and LHCb find that the asymmetry in the invariant mass bin containing the $Z$-pole is larger than in the adjacent bins, as predicted. Several beyond the Standard Model scenarios proposed for the top-quark forward-backward asymmetry, including a 100 GeV axigluon, are disfavored by this combination of SM predictions and measurements. On the other hand, modified $Zb\bar{b}$ couplings can explain the 2$\sigma$ discrepancy in the bottom-quark forward-backward asymmetry at LEP1, while being consistent with the results of CDF and LHCb. It is also shown that $t$-channel $W$ exchange makes a non-negligible contribution to the charm-quark charge asymmetry.
\end{abstract}

\maketitle


\section{Introduction}
For several years, measurements of, and related to, the forward-backward asymmetry in top-quark pair production ($A_{FB}^{t\bar{t}}$) at the Tevatron were consistently higher than the Standard Model (SM) predictions~\cite{Aaltonen:2011kc, Abazov:2011rq, Aaltonen:2012it, CDF:2013gna, Aaltonen:2014eva}. In particular, the measurement of the forward-backward asymmetry at high invariant mass ($M_{t\bar{t}} > 450$ GeV) was quoted to be 3.4 standard deviations above the next-to-leading order (NLO) QCD prediction~\cite{Aaltonen:2011kc}. This result sparked much work -- both theoretical and experimental, within and beyond the SM (BSM) -- trying to find the source of the discrepancy. See~\cite{Aguilar-Saavedra:2014kpa} for a review.

One proposal to help solve the problem was to measure the forward-backward asymmetry in bottom-quark pair production ($A_{FB}^{b\bar{b}}$)~\cite{ Bai:2011ed, Strassler:2011vr}, and the analogous charge asymmetry at the LHC ($A_{C}^{b\bar{b}}$)~\cite{Kahawala:2011sm}, see also~\cite{Sehgal:1987wi}. Such a measurement would likely be difficult due to dominance of gluon fusion initiated bottom pair production, which does not generate an asymmetry, over $q\bar{q}$ initiated production, the dominant top pair production mechanism. In addition, there were expected to be further complications due to $b$-tagging inefficiencies, and processes such as neutral $B$-meson oscillations and cascade decays spoiling the correlation between the charge of the decay products and the charge of the parent $b$-quark. However, the upside would be valuable information about the flavor structure of the source of the top asymmetry. Calculations of the bottom-quark $A_{FB}$ in the SM including some electroweak (EW) effects had been made both before~\cite{Kuhn:1998kw}, and after~\cite{Manohar:2012rs} these proposals. 

In~\cite{Grinstein:2013mia}, it was realized that a $Z$-boson decaying to $b\bar{b}$ would have significant consequences for the analysis of the bottom-quark asymmetry.\footnote{The top-quark is too heavy for an on-shell $Z$ to decay to $t\bar{t}$, so these effects are not nearly as strong in $A_{FB}^{t\bar{t}}$.} It was found that near the $Z$-pole, tree level exchanges of electroweak gauge bosons dominated the SM contribution to the $A_{FB}$. This is unlike the top asymmetry where NLO QCD is the leading contribution. Far enough above the $Z$-pole, NLO QCD does provide the leading SM contribution to the bottom asymmetry. 

The effects of various BSM scenarios on $A_{FB}^{b\bar{b}}$, and their relation to $A_{FB}^{t\bar{t}}$, were also investigated in~\cite{Grinstein:2013mia}.\footnote{See~\cite{Saha:2011wr, Drobnak:2012cz, Delaunay:2012kf, Ipek:2013zi} for additional studies of BSM effects on $A_{FB}^{b\bar{b}}$.} However, the $t\bar{t}$ asymmetry discrepancy has recently been resolved through a combination of theoretical and experimental work, namely results from D0 using the full Tevatron Run-2 dataset~\cite{Abazov:2013wxa, Abazov:2014oea, Abazov:2014cca}, and a SM prediction of $A_{FB}^{t\bar{t}}$ at next-to-next-to-leading order (NNLO) in QCD~\cite{Czakon:2014xsa}. Nevertheless, in the interim, measurements of $A_{FB}^{b\bar{b}}$ by the CDF~\cite{CDF:2014a, CDF:2014b} and D0 collaborations~\cite{Abazov:2014ysa, Hogan:2015sva} and of $A_{C}^{b\bar{b}}$ by the LHCb~\cite{LHCb:2013bka, Aaij:2014ywa} collaboration were made, and a new $\sim 3 \sigma$ discrepancy has appeared. 

The preliminary results of CDF at high invariant mass~\cite{CDF:2014a} are both consistent with zero, and agree with the SM predictions of~\cite{Grinstein:2013mia}. In addition, CDF was able to exclude a wide axigluon with a mass of 200 GeV as an explanation of $A_{FB}^{t\bar{t}}$, while not excluding an axigluon with a mass of 345 GeV. The preliminary results of CDF at low mass~\cite{CDF:2014b} agree with the SM predictions of~\cite{Grinstein:2013mia}, including a measured asymmetry in the bin containing the $Z$-pole that is larger than the asymmetry in the adjacent invariant mass bins.

LHCb updated their preliminary analysis of 7 TeV data~\cite{LHCb:2013bka} after Ref.~\cite{Grinstein:2013mia} came out to include an additional invariant mass bin centered on the $Z$-pole. In their published result, Ref.~\cite{Aaij:2014ywa}, they find that central value of $A_{C}^{b\bar{b}}$ in this $Z$-pole bin is the largest of the three bins in the analysis, similar to the prediction for $A_{FB}^{b\bar{b}}$ in~\cite{Grinstein:2013mia}. However, there was no dedicated SM prediction for this measurement at the time it was released.

D0 did an analysis~\cite{Abazov:2014ysa, Hogan:2015sva} of the forward-backward asymmetry in charged $B$-meson production, $A_{FB}\left(B^{\pm}\right)$, rather than a jet based analysis like CDF and LHCb. Clearly, there will be none of the charge tagging systematic issues discussed earlier in an analysis with charged mesons. Also, because D0 used an inclusive sample of $B^{\pm}$ they have more events and thus a smaller statistical uncertainty as well. However, this inclusiveness comes at the price of including far more gluon fusion initiated events, which dilute the already small asymmetry. D0's result is $A_{FB}\left(B^{\pm}\right) = [-0.24 \pm 0.41 (\text{stat.}) \pm 0.19 (\text{syst.})] \%$. This is consistent with zero. However, this is also 3.3 standard deviations below their SM prediction of $A^{SM}_{FB}\left(B^{\pm}\right) = [2.55 \pm 0.76] \%$, which was made using MC@NLO+Herwig~\cite{Frixione:2002ik, Corcella:2000bw}. D0 also finds that the measured asymmetry is lower than their SM prediction for all pseudorapidities, and for $p_T\left(B^{\pm}\right) = 9 - 30$ GeV.

D0 has also made a measurement of the forward-backward asymmetry in $\Lambda_b^0$ and $\overline{\Lambda}_b^0$ production~\cite{Abazov:2015vea}. This baryon asymmetry measurement is consistent with zero though it has large uncertainties, and central values that are large in magnitude. Note that the $\Lambda_b^0$ asymmetry is expected to be dominated by hadronic effects~\cite{Rosner:2014gta}. For this reason it is not considered further in this work. The analogous effect at the LHC, $\sigma(p p \to \Lambda_b^0 X) > \sigma(p p \to \overline{\Lambda}_b^0 X)$, is seen at high rapidity by CMS~\cite{Chatrchyan:2012xg}, and in the preliminary results of LHCb~\cite{LHCb:2012xja}.

The predictions made in~\cite{Grinstein:2013mia} for CDF are updated to include mixed EW-QCD corrections in an approximate way. These corrections are found to be small in magnitude, and CDF's measurements are found to be in good agreement with the SM predictions. The charge asymmetry at 7 TeV measured by LHCb is found to be in good agreement with Standard Model. It is also predicted that the charge asymmetry at 13 \& 14 TeV will be smaller than at 7 \& 8 TeV. The SM asymmetry is predicted to be very small for D0, which is consistent with what was measured. Several BSM models, including a 100 GeV axigluon model,  are ruled out by this combination of SM predictions and measurements. On the other hand, it is shown that the $Zb\bar{b}$ couplings can be modified to explain the anomalous bottom-quark forward-backward asymmetry at LEP1~\cite{Z-Pole}, while being consistent with the results of CDF and LHCb.

The rest of the paper is organized as follows. In Sec.~\ref{sec:sm}, the setup for the SM calculation of the forward-backward asymmetry is given. Predictions for the asymmetries measured by CDF, LHCb, and D0 (and for those to be measured by LHCb) are given in Sec.~\ref{sec:res}. Following that, Sec.~\ref{sec:dis} discusses the approximation used for the mixed EW-QCD corrections to the asymmetry, the charm-quark charge asymmetry, and implications for BSM scenarios. Finally, conclusions are given in Sec.~\ref{sec:con}.

\section{Standard Model Calculation}
\label{sec:sm}
The Standard Model contribution to the top-quark forward-backward asymmetry has been investigated extensively~\cite{Kuhn:1998kw, Manohar:2012rs, Czakon:2014xsa, Antunano:2007da, Almeida:2008ug, Ahrens:2011uf, Hollik:2011ps, Kuhn:2011ri, Campbell:2012uf, Skands:2012mm, Bernreuther:2012sx, Dittmaier:2007wz, Melnikov:2010iu, Kidonakis:2015ona}.\footnote{Note that the NLO QCD contribution to the forward-backward asymmetry in $q \bar{q} \to Q \bar{Q}$ is in a one-to-one correspondence with the previously known NLO QED contribution to the forward-backward asymmetry in $e^+ e ^- \to \mu^+ \mu^-$~\cite{Berends:1973fd}.} Both the forward-backward and the charge asymmetries can be defined as follows,
\begin{equation}
A_{FB,\, C} = \frac{\sigma\left(Y > 0\right) - \sigma\left(Y < 0\right)}{\sigma\left(Y > 0\right) + \sigma\left(Y < 0\right)},
\end{equation}
where $\sigma$ is the cross section in a given bin, and the observable $Y$ is used to determine whether an event is ``forwards'' or ``backwards.'' The partonic level observable most closely related to what CDF measures is the so-called rest frame forward-backward asymmetry, corresponding to $Y = y_b - y_{\bar{b}}$, while D0's measurement is closely related to the lab frame asymmetry, $Y = y_b$. Here $y_{b (\bar{b})}$ is the rapidity of the (anti-)bottom quark, $y = \ln((E + p_z)/(E - p_z))/2$, and $\hat{z}$ is the proton direction at the Tevatron. The observable for the charge asymmetry measured by LHCb is $Y = y_b^2 - y_{\bar{b}}^2$.

The SM contributes to the forward-backward asymmetry at various orders in perturbation theory, which can be written schematically as
\begin{align}
\label{eq:afb1}
A_{FB} &= \frac{N}{D}  \\
&= \frac{\alpha^2 \tilde{N}_0 + \alpha_s^3 N_1 + \alpha_s^2 \alpha \tilde{N}_1 + \alpha_s^4 N_2 + \cdots}{\alpha_s^2 D_0 +\alpha^2 \tilde{D}_0 + \alpha_s^3 D_1 + \alpha_s^2 \alpha \tilde{D}_1 + \cdots}. \nn
\end{align}
At the energy scales relevant for hadron colliders, Eq.~\eqref{eq:afb1} can expanded in powers of coupling constants
\begin{align}
\label{eq:afb2}
A_{FB} =\, &\alpha_s \frac{N_1}{D_0} + \frac{\alpha^2}{\alpha_s^2} \frac{\tilde{N}_0}{D_0} + \alpha \frac{\tilde{N}_1}{D_0} \\
&+ \alpha_s^2 \left(\frac{N_2}{D_0} - \frac{N_1 D_1}{D_0^2}\right) + \cdots. \nn
\end{align}
Eq.~\eqref{eq:afb2} is the definition of the forward-backward asymmetry we will use in our calculations, as is commonly done. The approach is the same as that of~\cite{Grinstein:2013mia}. We apply the analytic formulas in the literature for the $\mathcal{O}(\alpha_s)$~\cite{Kuhn:1998kw} and $\mathcal{O}(\alpha^2 / \alpha_s^2)$~\cite{Hollik:2011ps} terms to the case of the bottom asymmetry. The Cuhre integration routine from the Cuba library~\cite{Hahn:2004fe} is used for the leading order and virtual+soft corrections, and MadGraph5\_aMC@NLO~\cite{Alwall:2014hca} is used for the hard radiation. Unless otherwise stated, the set of parton distribution functions (PDF) used for all calculations is the NNPDF2.3QED NLO grid~\cite{Ball:2013hta, Hartland:2012ia} with $\alpha_s\left(M_Z\right) = 0.119$ and $m_b = 4.75$ GeV. The other numerical values used in this analysis are: $M_Z = 91.1876$ GeV, $\Gamma_Z = 2.4592$ GeV, $\alpha\left(M_Z\right) = 1/127.940$, and $\sin^2\theta_W \equiv s_W^2 = 0.23126$, which were taken from~\cite{Agashe:2014kda}. Cuts and binning are discussed in Sec.~\ref{sec:res}.

The tree-level EW contribution to the total cross section is included in the denominator of the computation of $A_{FB,\,C}$ when the invariant mass bin contains the $Z$-pole. Even though this term is formally higher-order than what we are considering, it is enhanced enough near $M_{b\bar{b}} \approx M_Z$ to warrant inclusion. The numerically small flavor excitation contribution to the asymmetry, $q g \to b \bar{b} q$, is neglected in all cases except for the inclusive D0 asymmetry. Similarly, $t$-channel $W$-exchange is neglected for all bottom asymmetries, but is analyzed in the context of a charm-quark asymmetry.

A formula for the $N_2$ term was not given in Ref.~\cite{Czakon:2014xsa}, presumably due to its complicated nature, so we will drop the entire $\mathcal{O}(\alpha_s^2)$ term. As was done before $N_2$ was known, to compensate for this neglect of higher-order corrections we assign an uncertainty to the calculation of 30\% of the $\mathcal{O}(\alpha_s)$ contribution, originating from $\alpha_s D_1 \approx 0.3 D_0$. Ref.~\cite{Czakon:2014xsa} finds that including $D_1$, but not $N_2$ in the NLO QCD calculation of the top $A_{FB}$ decreases the asymmetry by 25\%, and that the full $\mathcal{O}(\alpha_s^2)$ term increases the top asymmetry by 13\%.

In~\cite{Grinstein:2013mia} it was noted that the $\mathcal{O}(\alpha)$ contribution to $A_{FB}^{b\bar{b}}$ is small compared to the $\mathcal{O}(\alpha_s)$ and $\mathcal{O}(\alpha^2 / \alpha_s^2)$ terms, and their associated uncertainties. That statement is made quantitative in this work by computing the $\mathcal{O}(\alpha)$ contribution in an approximate way. The QED corrections are known to be in a one-to-one correspondence with the QCD asymmetry; one simply makes the following replacement for a given partonic channel:
\begin{equation}
\frac{\alpha_s}{2} \left(\frac{d_{abc}}{4}\right)^2 \to 3  \alpha Q_q Q_b.
\end{equation}
However, the mass of the $Z$ spoils this correspondence for the weak corrections. In this approximation, first one treats the $Z$ as massless. Then, including real $Z$ radiation, there is a one-to-one correspondence with the QCD asymmetry,
\begin{equation}
\label{eq:Z}
\frac{\alpha_s}{2} d_{abc}^2 \to 12 \alpha \left(T^3_q - 2 Q_q s_W^2\right)\left(T^3_b - 2 Q_b s_W^2\right) 
\end{equation}
This is what's done in e.g.~\cite{Kuhn:2011ri}, for the top asymmetry as $2 m_t \gg M_Z$. However, since~\cite{Grinstein:2013mia} showed that the $Z$-pole is important for the bottom asymmetry, this result is then multiplied by a correction factor that attempts to account for the resonance structure of the $Z$,
\begin{equation}
\label{eq:weight}
\frac{\int\! dM_{b\bar{b}}^2 \left(xf(x)\right)^2 \text{Re} \left[\left(1-\mu_Z^2 / M_{b\bar{b}}^2\right)^{-1}\right]}{\int\! dM_{b\bar{b}}^2 \left(xf(x)\right)^2}.
\end{equation}
In Eq.~\eqref{eq:weight}, $\mu_Z^2 = M_Z^2 - i \Gamma_Z M_Z$, and $xf(x)$ is $x$ times the PDF set for a given light quark flavor with $x = M_{b\bar{b}}^2 / s$. The integration is over the range of $M_{b\bar{b}}^2$ for a given bin. The form of this correction factor is justified a posteriori due to the smallness of the $\mathcal{O}\left(\alpha\right)$ terms; a different functional form won't affect the result for the total asymmetry very much. This is discussed further in Sec.~\ref{sec:dis}. Note that an exact computation of the $\mathcal{O}(\alpha)$ contribution to the $t\bar{t}$ asymmetry was made by~\cite{Hollik:2011ps}. Similarly, the $\mathcal{O}(\alpha)$ contribution to the $b\bar{b}$ asymmetry can be extracted from~\cite{Kuhn:2009nf}. In addition, shortly after this work was made public, a dedicated, exact computation for the $\mathcal{O}(\alpha)$ contribution to the $A_{C}^{b\bar{b}}$ was given in Ref.~\cite{Gauld:2015qha}.

\section{Results}
\label{sec:res}
%
In this section, parton-level SM predictions are given for the forward-backward and charge asymmetry. The cuts and binning are tailored to match the experimental analyses. In the tables below, the superscript uncertainty comes from running the renormalization and factorization scales, $\mu_R = \mu_F = \mu$, from the chosen central value of $M_Z$ up to $2M_Z$. Similarly, the subscript uncertainty is due running $\mu$ down to $M_Z/2$. Sometimes there is a partial cancellation between the scale uncertainty from the tree level EW contribution and the NLO QCD contribution. The QCD piece gets smaller as the scale increases because it is proportional to $\alpha_s$, whereas the EW piece gets bigger as $\mu$ increases because it scales as $1 / \alpha_s^2$. The additional, symmetric uncertainty in the last column is due to the neglect of higher-order QCD contributions. As previously stated, this uncertainty is taken to be 30\% of the pure QCD asymmetry, as is typically done.

As a check on the calculation, we have computed the NLO QCD contribution to the top-quark forward-backward asymmetry. We find $A_{FB}^{t\bar{t}} = 7.34\%$, $A_{FB}^{t\bar{t}}\left(M_{t\bar{t}} < 450\, \text{GeV}\right) = 5.36\%$, and $A_{FB}^{t\bar{t}}\left(M_{t\bar{t}} > 450\, \text{GeV}\right) = 10.59\%$, in good agreement with other determinations.

In what follows, the rest-frame and lab-frame forward-backward asymmetries are denoted by $A_{FB}^{b\bar{b}}$ and $A_{FB}$ respectively.

\subsection{CDF High Mass Analysis}
Results for the bottom-quark forward-backward asymmetry for the CDF high mass analysis~\cite{CDF:2014a} are given in Table~\ref{tab:cdfhigh} in percent. A cut has been placed on the rapidity of the bottom quarks, $|y_{b,\, \bar{b}}| \leq 1.1$.  No cuts have been made on the hard radiation. The results for two of three bins are in good agreement with CDF's measurements. In the $225 - 325$ GeV invariant mass bin, the SM prediction is just 1.3$\sigma$ above CDF's result.

\begin{table*}
\centering
 \begin{tabular}{c c c c c }
 Bin & $\mathcal{O}(\alpha^2 / \alpha_s^2)$ & $\mathcal{O}(\alpha_s)$ & $\mathcal{O}(\alpha)$ & $A^{b\bar{b}}_{FB}[\%]$  \\ \hline\hline
  $150 \leq M_{b\bar{b}}/\text{GeV} < 225$ & $0.15^{+0.04}_{-0.04}$ & $2.43_{+0.06}^{-0.05}$ & $-0.15_{-0.00}^{+0.00}$ & $2.43 \pm 0.73\,^{-0.01}_{+0.02}$ \\
  $225 \leq M_{b\bar{b}}/\text{GeV} < 325$ & $0.20^{+0.06}_{-0.05}$ & $4.72_{+0.21}^{-0.20}$ & $-0.31_{-0.01}^{+0.01}$ &  $4.61 \pm 1.38\,^{-0.13}_{+0.15}$ \\
        $325 \leq M_{b\bar{b}}/\text{GeV} $ & $0.28^{+0.07}_{-0.06}$ & $8.99_{+0.71}^{-0.62}$ & $-0.57_{-0.03}^{+0.04}$  & $8.70 \pm 2.61\,^{-0.51}_{+0.61}$ \\  \hline\hline
  \end{tabular}
  \caption{$A_{FB}^{b\bar{b}}$ in percent broken down by the various contributions and into different bins for the CDF high mass analysis. Here $|y_{b,\, \bar{b}}| \leq 1.1$, and no cuts have been placed on the hard radiation.}
  \label{tab:cdfhigh}
\end{table*}

The mixed EW-QCD corrections, $\mathcal{O}(\alpha)$, decrease the asymmetry. On the other hand, tree-level EW contribution, which was neglected in the evaluation of the high-mass asymmetry in~\cite{CDF:2014a}, is positive though smaller in magnitude than the mixed EW-QCD corrections. The uncertainty due to the neglect of higher-order QCD terms is larger than the magnitude of the sum of the two EW contributions, and the scale uncertainty is comparable in magnitude (or larger than) the total EW contribution. 

The astute reader will notice that the predictions for the $\mathcal{O}(\alpha_s)$ asymmetry in Tab.~\ref{tab:cdfhigh} are slightly larger than those in Table I of~\cite{CDF:2014a}, which used MSTW2008 NLO PDFs~\cite{Martin:2009iq}. Indeed, though the two calculations agree within the scale uncertainty, the choice of PDF set seems to make a small difference in the prediction for the central value of the asymmetry.\footnote{Using LO versus NLO PDFs will also lead to different predictions, as can be seen in e.g.~\cite{Gauld:2014pxa}.} For example, using Hollik and Pagani's Eq.~(7)~\cite{Hollik:2011ps}, we find using NNPDF2.3QED NLO PDFs that the $\mathcal{O}\left(\alpha^2 / \alpha_s^2\right)$ contributions to $A_{FB}^{t\bar{t}}$ are $2.5 \cdot 10^{-3}$ and $3.6 \cdot 10^{-4}$ in the $u\bar{u}$ and $d\bar{d}$ channels respectively. Using MSTW2008 NLO PDFs, we instead find $\mathcal{O}(\alpha^2 / \alpha_s^2)_{u\bar{u}} = 1.8 \cdot 10^{-3}$ and $\mathcal{O}(\alpha^2 / \alpha_s^2)_{d\bar{d}} = 3.2 \cdot 10^{-4}$, slightly smaller than the previous evaluation. Hollik and Pagani themselves find $2.3 \cdot 10^{-3}$ and $3 \cdot 10^{-4}$ in the $u\bar{u}$ and $d\bar{d}$ channels respectively~\cite{Hollik:2011ps} using MRST2001 LO PDFs~\cite{Martin:2002dr}.

\subsection{CDF Low Mass Analysis}
Results for the bottom-quark forward-backward asymmetry for the CDF low mass analysis~\cite{CDF:2014b} are given in Table~\ref{tab:cdflow} in percent. The cuts $p_{T b,\, \bar{b}} > 15$ GeV and $M_{b\bar{b}} > 35$ GeV were placed on all of the bins. In Ref.~\cite{Grinstein:2013mia}, only the $p_T$ cut had been implemented.\footnote{The label of the bin in the second row, ``$35 \leq M_{b\bar{b}} / \text{GeV} < 75$,'' of Table I in~\cite{Grinstein:2013mia} is misleading because there wasn't actually a $M_{b\bar{b}} > 35$ GeV cut on this bin.} The cut on the rapidity of the bottom quarks, $|y_{b,\, \bar{b}}| \leq 1$, is the same as in~\cite{Grinstein:2013mia}. In this work, two cuts were placed on hard radiation instead of one. In particular, the opening angle, $\Delta \phi$, between the bottom and anti-bottom in the plane transverse to the beam line is required to be greater than 2.8 radians.\footnote{In~\cite{Grinstein:2013mia}, this was referred to as the ``acollinearity'' cut.} In addition, to match CDF's cuts, the bottoms are also required to satisfy,
\begin{equation}
\label{eq:pt}
|p_{T b} - p_{T \bar{b}}| < 0.6\, \text{max}\{p_{T b},\, p_{T \bar{b}}\}.
\end{equation}
The cut in Eq.~\eqref{eq:pt} was not included in the analysis of Ref.~\cite{Grinstein:2013mia}.

\begin{table*}
\centering
 \begin{tabular}{  c  c c  c c }
 Bin & $\mathcal{O}(\alpha^2 / \alpha_s^2)$ & $\mathcal{O}(\alpha_s)$ & $\mathcal{O}(\alpha)$ & $A^{b\bar{b}}_{FB}[\%]$  \\ \hline\hline
  $35 \leq M_{b\bar{b}}/\text{GeV} < 75$ & $0.00^{+0.00}_{-0.00}$ & $0.20_{+0.01}^{-0.01}$ & $-0.01$ & $0.19 \pm 0.06\,^{-0.01}_{+0.01}$ \\
  $75 \leq M_{b\bar{b}}/\text{GeV} < 95$ & $2.01^{+0.54}_{-0.47}$ & $0.52_{+0.03}^{-0.02}$ & $-0.05$ &  $2.49 \pm 0.16\,^{+0.52}_{-0.44}$ \\
$95 \leq M_{b\bar{b}}/\text{GeV} < 130$ & $0.56^{+0.17}_{-0.14}$ & $0.89_{+0.02}^{-0.02}$ & $-0.01$ &  $1.44 \pm 0.27\,^{+0.16}_{-0.12}$ \\
        $130 \leq M_{b\bar{b}}/\text{GeV} $ & $0.15^{+0.05}_{-0.04}$ & $2.11_{+0.08}^{-0.07}$ & $-0.13$  & $2.14 \pm 0.63\,^{-0.01}_{+0.03}$ \\ \hline
      $0.0 \leq |\Delta y_{b\bar{b}}| < 0.5$ & $0.05^{+0.01}_{-0.01}$ & $0.13_{+0.01}^{-0.01}$ & $-0.00$  & $0.18 \pm 0.04\,^{+0.00}_{-0.00}$ \\
      $0.5 \leq |\Delta y_{b\bar{b}}| < 1.0$ & $0.11^{+0.02}_{-0.02}$ & $0.29_{+0.02}^{-0.02}$ & $-0.01$ & $0.38 \pm 0.09\,^{+0.01}_{-0.00}$ \\
   $1.0 \leq |\Delta y_{b\bar{b}}| \leq 2.0$ & $0.13^{+0.03}_{-0.03}$ & $0.39_{+0.03}^{-0.02}$ & $-0.02$  & $0.51 \pm 0.12\,^{+0.01}_{-0.00}$ \\ \hline
$35 \leq M_{b\bar{b}}/\text{GeV}$             & $0.09^{+0.02}_{-0.02}$ & $0.25_{+0.02}^{-0.01}$ & $-0.01$ &  $0.34 \pm 0.08\,^{+0.01}_{-0.00}$ \\ \hline\hline
  \end{tabular}
  \caption{$A_{FB}^{b\bar{b}}$ in percent broken down by the various contributions and into different bins for the CDF low mass analysis. No scale uncertainties are given for the $\mathcal{O}(\alpha)$ terms as they are smaller than $10^{-4}$. See the text for a discussion of the cuts used in this analysis.}
  \label{tab:cdflow}
\end{table*}

No scale uncertainties are given for the $\mathcal{O}(\alpha)$ terms as they are smaller than $10^{-4}$, which is due to the fact that the mixed EW-QCD corrections are themselves rather small. The $\mathcal{O}(\alpha^2 / \alpha_s^2)$ terms are all either in good agreement with the findings of~\cite{Grinstein:2013mia} or are slightly higher. However, the bins that are slightly higher are exactly the bins that we expect to be affected by now having both a $p_T$ and an $M_{b\bar{b}}$ cut. The affected bins are the $35 \leq M_{b\bar{b}}/\text{GeV} < 75$ bin, all three rapidity bins, and the $35 \leq M_{b\bar{b}}/\text{GeV}$ bin. For the $\mathcal{O}(\alpha_s)$ terms, the cuts on the hard radiation increase the asymmetry by $8-27\%$, depending on the bin. As expected, without cuts on hard radiation, the asymmetries given here and in Ref.~\cite{Grinstein:2013mia} are in good agreement. 

\subsection{LHCb}
Results for the bottom-quark charge asymmetry for the LHCb 7 TeV analysis~\cite{Aaij:2014ywa} are given in Table~\ref{tab:lhcb7}. The cuts, $2 \leq y_{b,\, \bar{b}} \leq 4$ and $E_{T\, b,\, \bar{b}} > 20$ GeV, have been applied. In addition, the bottoms are required to have an opening angle in plane transverse to the beam line satisfying $\Delta \phi > 2.6$ rad. The prediction in each bin for $\sqrt{s} = 7$ TeV is in good agreement with measured value. Without the inclusion of the tree-level EW contribution to the total cross section, the prediction for $A^{b\bar{b}}_{C}\left(75 \leq M_{b\bar{b}}/\text{GeV} < 105\right)$ increases from 3.50\% to 3.81\%. 

\begin{table*}
\centering
 \begin{tabular}{  c  c c c }
 Bin & $\mathcal{O}(\alpha^2 / \alpha_s^2)$ & $\mathcal{O}(\alpha_s)$ &  $A^{b\bar{b}}_{FB}[\%]$  \\ \hline\hline
  $40 \leq M_{b\bar{b}}/\text{GeV} < 75  $ & $0.00^{+0.00}_{-0.00}$ & $0.46_{+0.04}^{-0.03}$ & $0.46 \pm 0.14\,^{-0.03}_{+0.04}$ \\
  $75 \leq M_{b\bar{b}}/\text{GeV} < 105$ & $2.48^{+0.59}_{-0.52}$ & $1.02_{+0.08}^{-0.07}$ & $3.50 \pm 0.31\,^{+0.52}_{-0.43}$ \\
$105 \leq M_{b\bar{b}}/\text{GeV}           $ & $0.25^{+0.07}_{-0.06}$ & $1.53_{+0.09}^{-0.06}$ & $1.79 \pm 0.46\,^{+0.01}_{+0.03}$ \\
\hline\hline
  \end{tabular}
  \caption{$A_{C}^{b\bar{b}}$ in percent for LHCb with $\sqrt{s} = 7$ TeV. Here $2 \leq y_{b,\, \bar{b}} \leq 4$, $E_{T b,\, \bar{b}} > 20$ GeV, and $\Delta \phi > 2.6$ rad.}
  \label{tab:lhcb7}
\end{table*}

Predictions have also been made for the bottom charge asymmetry at LHCb for $\sqrt{s} = 8,\, 13,\, \text{and}\, 14$ TeV. There are two differences between these calculations and the 7 TeV analysis. First, the rapidity cut has been changed to $2.2 \leq y_{b,\, \bar{b}} \leq 4.2$, and second, a low $p_T$ bin has been included. Results for the SM charge asymmetry at $\sqrt{s} = 8,\, 13,\, \text{and}\, 14$ TeV are given in Tables~\ref{tab:lhcb8},~\ref{tab:lhcb13}, and~\ref{tab:lhcb14} respectively. Due to the increase in gluon fusion initiated bottom production at higher $s$ (smaller $x$ for a given $M_{b\bar{b}}$), the asymmetry is predicted to be smaller at 13 \& 14 TeV than it is at 7 \& 8 TeV.

\begin{table*}
\centering
 \begin{tabular}{  c  c c c }
 Bin & $\mathcal{O}(\alpha^2 / \alpha_s^2)$ & $\mathcal{O}(\alpha_s)$ &  $A^{b\bar{b}}_{C}[\%]$  \\ \hline\hline
  $10 \leq p_{T b,\, \bar{b}}/\text{GeV} \leq 20$ & $0.00^{+0.00}_{-0.00}$ & $0.19_{+0.02}^{-0.01}$ & $0.19 \pm 0.06\,^{-0.01}_{+0.02}$ \\
  $40 \leq M_{b\bar{b}}/\text{GeV} < 75  $          & $0.00^{+0.00}_{-0.00}$ & $0.48_{+0.03}^{-0.03}$ & $0.48 \pm 0.14\,^{-0.03}_{+0.04}$ \\
  $75 \leq M_{b\bar{b}}/\text{GeV} < 105$          & $2.56^{+0.59}_{-0.52}$ & $0.75_{+0.07}^{-0.05}$ & $3.31 \pm 0.22\,^{+0.53}_{-0.45}$ \\
$105 \leq M_{b\bar{b}}/\text{GeV}           $          & $0.27^{+0.07}_{-0.06}$ & $1.54_{+0.10}^{-0.08}$ & $1.81 \pm 0.46\,^{-0.00}_{+0.04}$ \\
\hline\hline
  \end{tabular}
  \caption{$A_{C}^{b\bar{b}}$ in percent for LHCb with $\sqrt{s} = 8$ TeV. Here $2.2 \leq y_{b,\, \bar{b}} \leq 4.2$, $E_{T b,\, \bar{b}} > 20$ GeV, and $\Delta \phi > 2.6$ rad.}
  \label{tab:lhcb8}
\end{table*}

\begin{table*}
\centering
 \begin{tabular}{  c  c c c }
 Bin & $\mathcal{O}(\alpha^2 / \alpha_s^2)$ & $\mathcal{O}(\alpha_s)$ &  $A^{b\bar{b}}_{C}[\%]$  \\ \hline\hline
  $10 \leq p_{T b,\, \bar{b}}/\text{GeV} \leq 20$ & $0.00^{+0.00}_{-0.00}$ & $0.10^{-0.01}_{+0.01}$ & $0.10 \pm 0.03\,_{+0.01}^{-0.01}$ \\
  $40 \leq M_{b\bar{b}}/\text{GeV} < 75  $         & $0.00^{+0.00}_{-0.00}$ & $0.17^{-0.01}_{+0.02}$ & $0.17 \pm 0.05\,_{+0.02}^{-0.01}$ \\
  $75 \leq M_{b\bar{b}}/\text{GeV} < 105$         & $1.47^{+0.35}_{-0.30}$ & $0.44^{-0.03}_{+0.04}$ & $1.91 \pm 0.13\,_{-0.26}^{+0.32}$ \\
$105 \leq M_{b\bar{b}}/\text{GeV}           $         & $0.14^{+0.04}_{-0.03}$ & $0.70^{-0.04}_{+0.05}$ & $0.84 \pm 0.21\,_{+0.01}^{+0.00}$ \\
\hline\hline
  \end{tabular}
  \caption{Same as Table~\ref{tab:lhcb8}, but for $\sqrt{s} = 13$ TeV.}
  \label{tab:lhcb13}
\end{table*}

\begin{table*}
\centering
 \begin{tabular}{  c  c c c }
 Bin & $\mathcal{O}(\alpha^2 / \alpha_s^2)$ & $\mathcal{O}(\alpha_s)$ &  $A^{b\bar{b}}_{C}[\%]$  \\ \hline\hline
  $10 \leq p_{T b,\, \bar{b}}/\text{GeV} \leq 20$ & $0.00^{+0.00}_{-0.00}$ & $0.08_{+0.01}^{-0.01}$ & $0.08 \pm 0.02\,^{-0.01}_{+0.01}$ \\
  $40 \leq M_{b\bar{b}}/\text{GeV} < 75  $         & $0.00^{+0.00}_{-0.00}$ & $0.21_{+0.02}^{-0.01}$ & $0.21 \pm 0.06\,^{-0.01}_{+0.02}$ \\
  $75 \leq M_{b\bar{b}}/\text{GeV} < 105$         & $1.34^{+0.32}_{-0.24}$ & $0.37_{+0.03}^{-0.02}$ & $1.72 \pm 0.11\,^{+0.30}_{-0.24}$ \\
$105 \leq M_{b\bar{b}}/\text{GeV}           $         & $0.13^{+0.06}_{-0.04}$ & $0.84_{+0.06}^{-0.04}$ & $0.97 \pm 0.25\,^{-0.01}_{+0.03}$ \\
\hline\hline
  \end{tabular}
  \caption{Same as Table~\ref{tab:lhcb8}, but for $\sqrt{s} = 14$ TeV.}
  \label{tab:lhcb14}
\end{table*}

Unlike the case of the CDF low mass analysis, the $\Delta \phi$ cut makes very little difference for the LHCb analysis. It only increases the $\mathcal{O}\left(\alpha_s\right)$ asymmetry by $1-2\%$. The scale uncertainty is artificially small in the $M_{b\bar{b}} > 105$ GeV bin due to a partial cancellation between the $\mathcal{O}(\alpha^2 / \alpha_s^2)$ and $\mathcal{O}(\alpha_s)$ terms. In fact, a peculiar feature of this cancellation is that for $\sqrt{s} = 7\, \text{and}\, 13$ TeV, the asymmetry is the smallest in the $M_{b\bar{b}} > 105$ GeV bin for $\mu = M_Z$ rather than $\mu = 2 M_Z$ or $M_Z / 2$.\footnote{For Fig.~\ref{fig:cdflow} (lower left panel), the $\mu = 2 M_Z$ prediction for the SM asymmetry is used for the central value.}

\subsection{D0}
\label{sec:resd0}
The same perturbative calculation used for the CDF and LHCb results is again used for the D0 analysis.  We are mindful that because D0 measured an exclusive hadronic final state, a perturbative calculation may not necessarily be relevant. However, D0 finds the that rms width of the distribution of $\left(\eta_b - \eta_B\right)$ is 0.11~\cite{Abazov:2014ysa}; see~\cite{Hogan:2015sva} for more information. This suggests that hadronization does not significantly affect $A_{FB}$ in $B^{\pm}$ meson production.

The following cuts are made: $0.1 \leq |y_b| \leq 2.1$ to match D0's analysis, and $|y_{\bar{b}}| < 2.1$ to simplify the calculation. Note that D0 uses pseudorapidity in its analysis, whereas rapidity is used in this calculation. 

We find that the $\mathcal{O}\left( \alpha_s\right)$ contribution to the inclusive asymmetry is $9.8 \pm 0.3^{-0.1}_{+0.1} \cdot 10^{-5}$. The tree-level EW contribution to the asymmetry is $4.8^{+0.7}_{-0.6} \cdot 10^{-6}$. Since this asymmetry is so small, the flavor excitation piece is also considered. A quick calculation using MadGraph~\cite{Alwall:2014hca} gives $\mathcal{O}\left( \alpha_s\right)_{qg} = 1.1 \pm 0.7(\text{stat.}) \cdot 10^{-4}$, the same order of magnitude as the $q\bar{q}$ initiated asymmetry. Combining these results yields our final prediction for the inclusive asymmetry, $A_{FB} = \left(2.1 \pm 0.8\right) \cdot 10^{-4}$. A very small asymmetry is exactly what's expected from a sample of bottom quarks without cuts on $p_{T}$ or $M_{b\bar{b}}$. There is nothing to suppress the symmetric gluon fusion production process, so the asymmetry is diluted away to almost nothing. D0's result for the inclusive asymmetry is $A_{FB}\left(B^{\pm}\right) = [-0.24 \pm 0.41 (\text{stat.}) \pm 0.19 (\text{syst.})] \%$~\cite{Abazov:2014ysa}, which is in good agreement with this calculation. 

\begin{figure*}
  \centering
 \subfloat{\label{fig:d0eta}\includegraphics[width=0.5\textwidth]{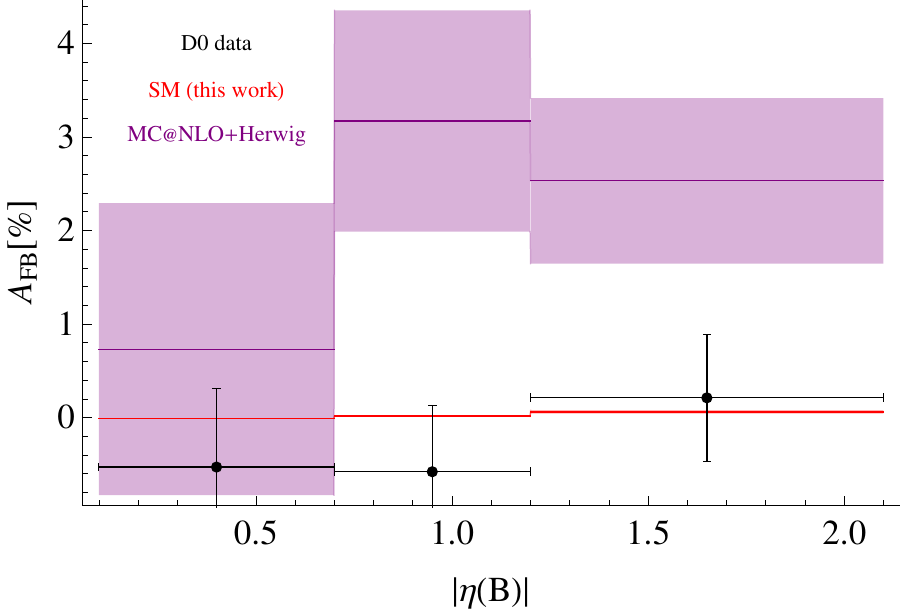}} 
  \subfloat{\label{fig:d0pt}\includegraphics[width=0.5\textwidth]{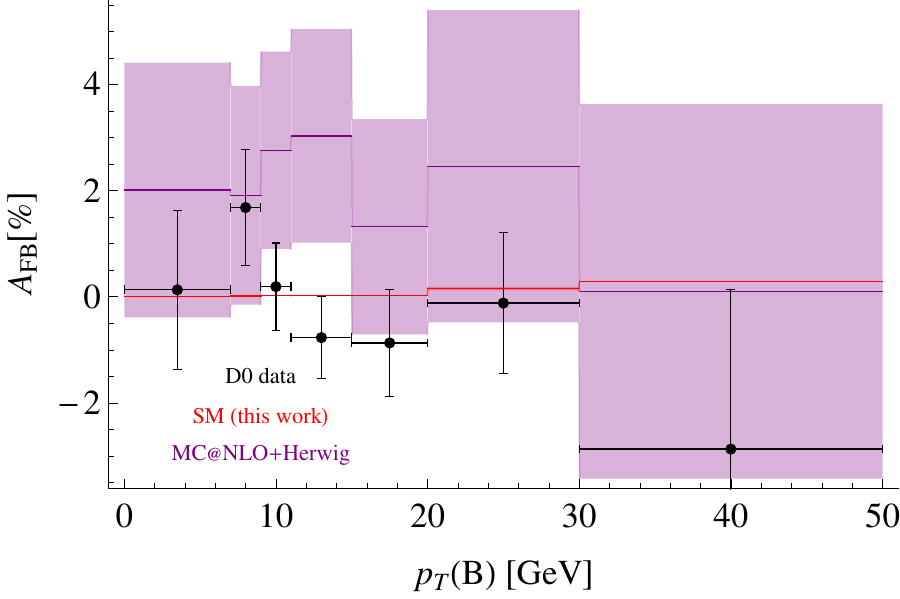}} 
   \caption{$A_{FB}$ vs. $|\eta(B)|$ (left) and $p_T(B)$ (right). In both cases, data from D0 and their corresponding predictions~\cite{Abazov:2014ysa}, which were made using MC@NLO+Herwig, are shown in black and purple respectively. The SM predictions from this work are in red.}
 \label{fig:d0}
   \end{figure*}

Predictions are also made for the $|\eta(B)|$ and $p_T(B)$ distributions of the asymmetry measured by D0, which are given in Fig.~\ref{fig:d0}. $A_{FB}$ is given as a function of $|\eta(B)|$ in the left panel and of $p_T(B)$ in the right panel. Data from D0 and their corresponding predictions~\cite{Abazov:2014ysa}, which were made using MC@NLO+Herwig are shown in black and purple respectively. The SM predictions from this work are in red. The absolute value of the rapidity and the $p_T$ of the bottom-quark are used for $|\eta(B)|$ and $p_T(B)$ respectively. The flavor excitation process is not included in these distributions. Only the uncertainty due to the neglect of higher-order terms is included; no scale uncertainty is calculated for these distributions. The  SM calculation is consistent with D0's measurements in all bins except for the $7 \leq p_T / \text{GeV} < 9$ bin, where the asymmetry is measured to be larger than what is predicted.

As previously noted, there is good agreement between the SM predictions in this work and the D0 measurements. On the other hand, the D0 observations and their predictions from MC@NLO+Herwig differ at the 3$\sigma$ level, with the MC prediction being larger than what was measured the majority of the time. Note also that the D0 baryon analysis also finds that MC@NLO+Herwig predicts an asymmetry that is larger than what was measured~\cite{Abazov:2015vea}.

\section{Discussion}
\label{sec:dis}
In what follows a discussion is given of the estimation of the mixed EW-QCD corrections, the charm-quark charge asymmetry, and the implications of this work for BSM scenarios.

\subsection{Estimate of Mixed EW-QCD Corrections}
Inspecting Tables~\ref{tab:cdfhigh} and~\ref{tab:cdflow} it is seen that the $\mathcal{O}(\alpha)$ terms make small contributions to the bottom $A_{FB}$ relative to the $\mathcal{O}(\alpha_s)$ and $\mathcal{O}(\alpha^2 / \alpha_s^2)$ terms, and their associated uncertainties. This smallness justifies the functional form of the approximation used in Eq.~\eqref{eq:weight} a posteriori, simply because changing the functional form won't make much of a difference in the total prediction for the asymmetry. Case in point, consider the $95 \leq M_{b\bar{b}}/\text{GeV} < 130$ bin in Table~\ref{tab:cdflow}, where there is a partial cancellation between the QED and the weak contribution, with $\mathcal{O}(\alpha)_{\text{QED}} \approx -0.05\%$ and $\mathcal{O}(\alpha)_{\text{weak}} \approx+0.04\%$. Now consider what would happen if the following form for the weight was used,
\begin{equation}
\label{eq:log}
\frac{\hat{s}}{\hat{s} - M_Z^2} \ln\left(1 - \frac{\hat{s}}{M_Z^2}\right),
\end{equation}
with $\hat{s}$ being the integration variable.\footnote{This is an example of the type of logarithms that were resummed to make predictions for $A_{FB}^{b\bar{b}}$ in Ref~\cite{Manohar:2012rs}.} Such terms arise from interference between tree-level gluon exchange and box diagrams containing one $Z$ and one gluon, see e.g.~\cite{Kuhn:2009nf}. Replacing $M_Z$ with $\mu_Z$ and taking the real part, the effect of the $\log$ is to change the sign of $\mathcal{O}(\alpha)_{\text{weak}}$, so that $\mathcal{O}(\alpha) \approx - 0.05\% - 0.04\% = - 0.09\%$.\footnote{As can be seen in~\cite{Kuhn:2009nf}, the $\cos\theta_{\text{CM}}$ dependence of the log and non-log terms are numerically similar after integration so that $\left|\mathcal{O}(\alpha)_{\text{weak}}\right| \approx 0.04\%$ in both cases.}  Considering the $\mathcal{O}(\alpha)$ term by itself, this looks like a large effect, almost an order of magnitude increase. However, the sum of all the contributions to the asymmetry in the $95 \leq M_{b\bar{b}}/\text{GeV} < 130$ bin only changes from $1.44\%$ to $1.34\%$. This change is smaller than the scale uncertainty, and it is also smaller than the uncertainty due to the neglect of higher-order QCD terms.

\subsection{Charm-Quark Charge Asymmetry}
There is also some interest in measuring the charge asymmetry in charm-quark production~\cite{Strassler:2011vr, Kahawala:2011sm}, and LHCb may well have the charm-tagging capabilities to do so. A full study of the SM contribution to the charm asymmetry is left for future work. Instead, in this work, we consider a previously neglected, tree-level contribution to the heavy quark charge asymmetry due to $t$-channel $W$ exchange. For the top and bottom asymmetries, this contribution is rightfully neglected because it is highly Cabibbo-Kobayashi-Maskawa (CKM) (or bottom PDF) suppressed. However, there is less suppression for the charm-quark asymmetry because it is only a first to second generation transition. To the best of our knowledge, the asymmetric piece of the interference between $s$-channel gluon exchange and $t$-channel $W$ exchange has not previously been given in the literature,
\begin{widetext}
\begin{equation}
\frac{d \sigma_A^{\alpha_s \alpha}}{d \cos \theta} = - \frac{\alpha_s \alpha \left| V_{Qq} \right|^2}{\sin^2\theta_W} \frac{C_F}{N_C} \frac{\pi \beta}{8 s} \frac{c}{\rho^2} \frac{2 \rho^2 \left(3 + c^2 + 4 \rho^2\right) - m^2 \left(1 - c^2 + 4 \rho^2\right) + 8 m^4}{\left(1 + 2 \rho^2 - 2 m^2\right)^2 - c^2},
\end{equation}
\end{widetext}
where $m^2 = m_Q^2 / s,\, \rho^2 = M_W^2 / s,\, \beta = \sqrt{1- 4 m^2},\, c = \beta \cos \theta$, and $V$ is the CKM matrix. The square of the $t$-channel diagram is suppressed by $(\alpha / \alpha_s)(|V_{cd}|^2/ s_W^2) \approx 1\%$ relative to the interference term. The amplitude for $d \bar{s}$ (and $s \bar{d}$) initiated $c\bar{c}$ production has less CKM suppression than the $d\bar{d}$ process. However, the $d \bar{s} + s \bar{d}$ initiated $t$-channel $W$ exchange does not interfere with $s$-channel gluon exchange, so its effect on the asymmetry should also be small compared to the inference contribution.

Potential collinear singularities are regulated by the mass of the $W$, so based on counting powers of coupling constants, the size of the asymmetry due to $W$ exchange should be suppressed by about an order of magnitude compared to the NLO QCD contribution to the asymmetry. We find the $\mathcal{O}\left(\alpha_s \alpha\right)$ contribution to $A_{FB}^{c\bar{c}}$ is $-0.4\%$ for $350 \leq M_{c\bar{c}} / \text{GeV} \leq 950$ and $\left|y_{c,\bar{c}}\right| \leq 1.84$ with $M_W = 80.385$ GeV and $m_c =\sqrt{2}$ GeV. Manohar and Trott find the NLO QCD contribution to $A_{FB}^{c\bar{c}}$ is $6.7\%$ for $350 \leq M_{c\bar{c}} / \text{GeV} \leq 650$ and $18\%$ for $650 \leq M_{c\bar{c}} / \text{GeV} \leq 950$~\cite{Manohar:2012rs}. This is consistent with the naive scaling estimate; $t$-channel $W$ exchange is small, but not negligible for the charm asymmetry. On the other hand, we find the $\mathcal{O}\left(\alpha_s \alpha\right)$ contribution to $A_{FB}^{b\bar{b}}$ for the same invariant mass range to be $-6 \cdot 10^{-6}$, which is completely negligible.

\subsection{Implications for BSM Scenarios}
Two different BSM scenarios are investigated in this work. First, as was done in~\cite{Grinstein:2013mia}, the effect on $A_{FB}^{b\bar{b}}$ of BSM physics models proposed for $A_{FB}^{t\bar{t}}$  is considered. However, this analysis comes with the caveat that since the anomalous top-quark forward-backward asymmetry has been resolved, none of the following models are necessarily viable anymore. Second, constraints on modified $Zb\bar{b}$ couplings are derived from measurements of $A_{FB,\, C}^{b\bar{b}}$ near the $Z$-pole by CDF and LHCb. A comparison of these bounds with the analogous results from LEP1~\cite{Z-Pole} is given as well.

The forward-backward asymmetry as a function of invariant mass is plotted in Fig.~\ref{fig:cdflow}. CDF's low mass measurement~\cite{CDF:2014b} are in the top panel. The LHCb 7 TeV data is on the bottom left, and the CDF high mass data is on the bottom right. The SM predictions from this work are shown in red. Plotted in blue in Fig.~\ref{fig:cdflow} is a prediction from the axigluon model~\cite{Frampton:1987dn}. The parameters used in the plot are, $M_{G^{\prime}} = 100$ GeV, $\Gamma_{G^{\prime}} = M_{G^{\prime}} / 10$, and $g_a = 0.476$. These parameters were taken from Ref.~\cite{Gross:2012bz}, which throughly investigate the bounds on the axigluon models that were relevant as BSM explanations of $A_{FB}^{t\bar{t}}$. The other parameter choices in Ref.~\cite{Gross:2012bz} do not cause significant deviations from the SM, as can be seen in Fig. 1 of~\cite{Grinstein:2013mia}. The measurements by CDF~\cite{CDF:2014b} combined with the SM predictions in this work disfavors the 100 GeV axigluon. 

Furthermore, both of the benchmark points from~\cite{Grinstein:2013mia} for the scalar weak doublet model~\cite{Blum:2011fa} are also disfavored. This can be seen in Fig.~\ref{fig:cdflow4} where CDF's measurements~\cite{CDF:2014b} are in black, and the SM predictions from Table~\ref{tab:cdflow} are shown in red.  In addition, the 100 and 150 GeV flavor octet, electroweak triplet vectors of~\cite{Grinstein:2011yv, Grinstein:2011dz} are also disfavored, but 250 GeV vectors are consistent with this analysis. Predictions for the bottom asymmetry due to a 105 GeV scalar weak doublet and a flavor octet of 150 GeV EW triplet vectors are plotted in Fig.~\ref{fig:cdflow4} in brown and green respectively. The BSM contributions to $A_{FB}^{b\bar{b}}$ are computed using MadGraph~\cite{Alwall:2014hca}, and a statistical uncertainty of $0.3\%$ is included in addition to the SM uncertainties. 

\begin{figure*}
  \centering
 \subfloat{\label{fig:cdflow}\includegraphics[width=0.7\textwidth]{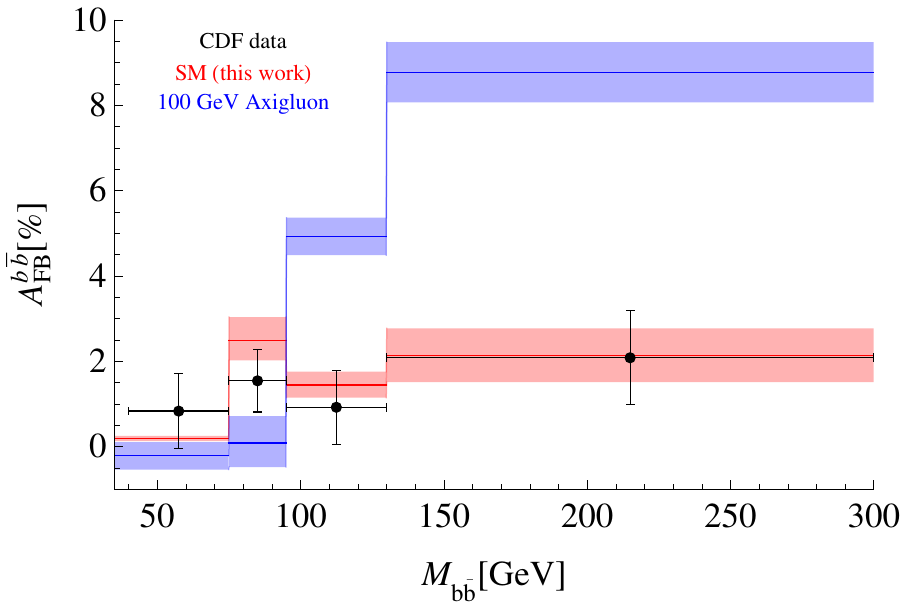}} \\
 \subfloat{\label{fig:lhcb}\includegraphics[width=0.5\textwidth]{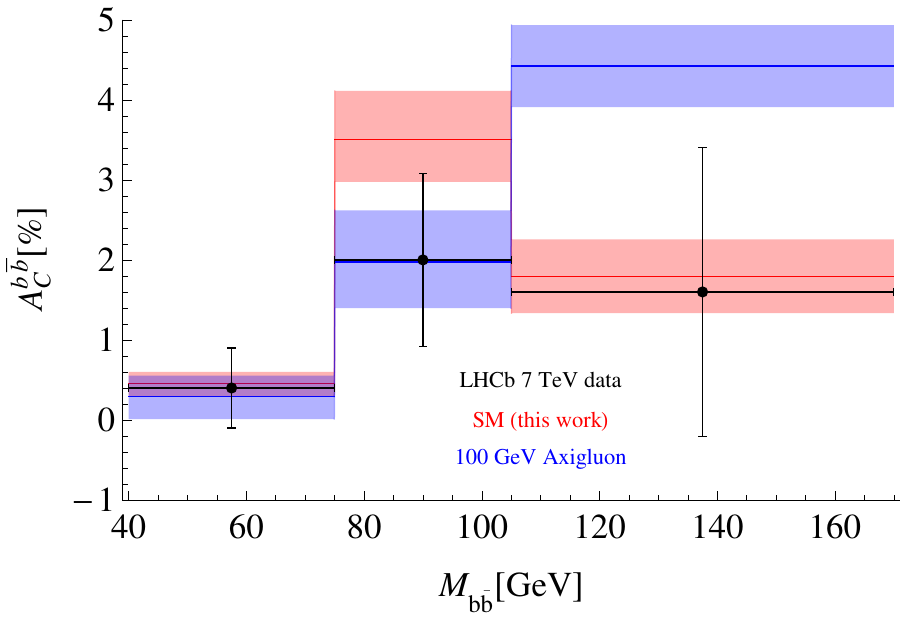}} 
  \subfloat{\label{fig:cdfhigh}\includegraphics[width=0.5\textwidth]{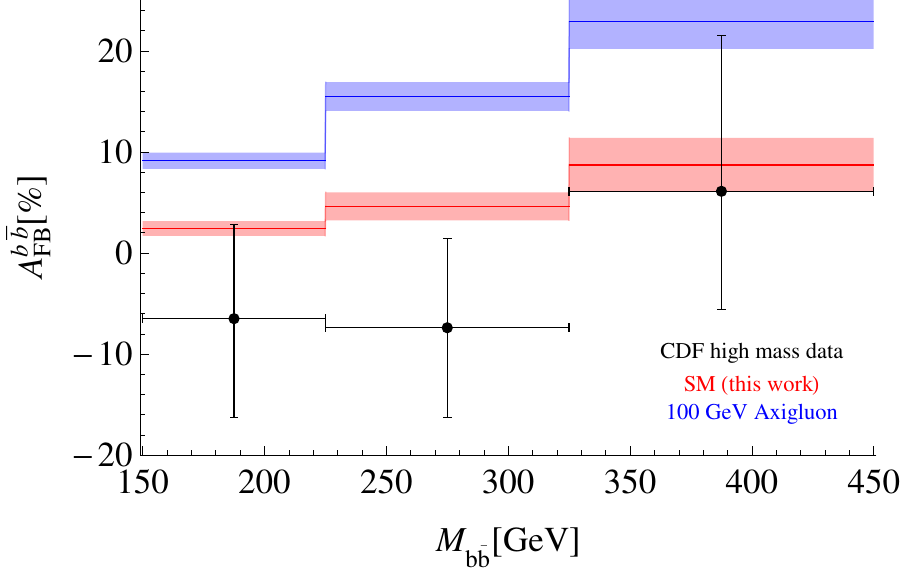}} 
   \caption{Top: CDF's low mass measurements of $A_{FB}^{b\bar{b}}$~\cite{CDF:2014b}. Bottom left: LHCb's 7 TeV measurements of $A_{C}^{b\bar{b}}$~\cite{Aaij:2014ywa} Bottom right: CDF's high mass measurements of $A_{FB}^{b\bar{b}}$~\cite{CDF:2014a}. The experimental data is shown in black. In all three plots, the SM predictions from this work are shown in red, and plotted in blue are predictions for the bottom asymmetry due to a 100 GeV axigluon.}
 \label{fig:cdflow}
   \end{figure*}

\begin{figure*}
  \centering
 \subfloat{\label{fig:SWD}\includegraphics[width=0.5\textwidth]{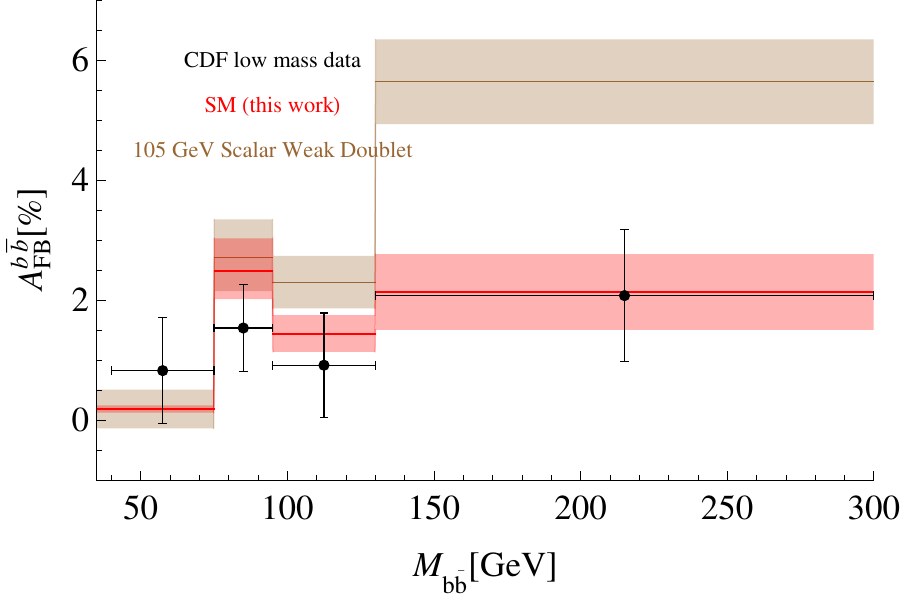}} 
  \subfloat{\label{fig:EWT}\includegraphics[width=0.5\textwidth]{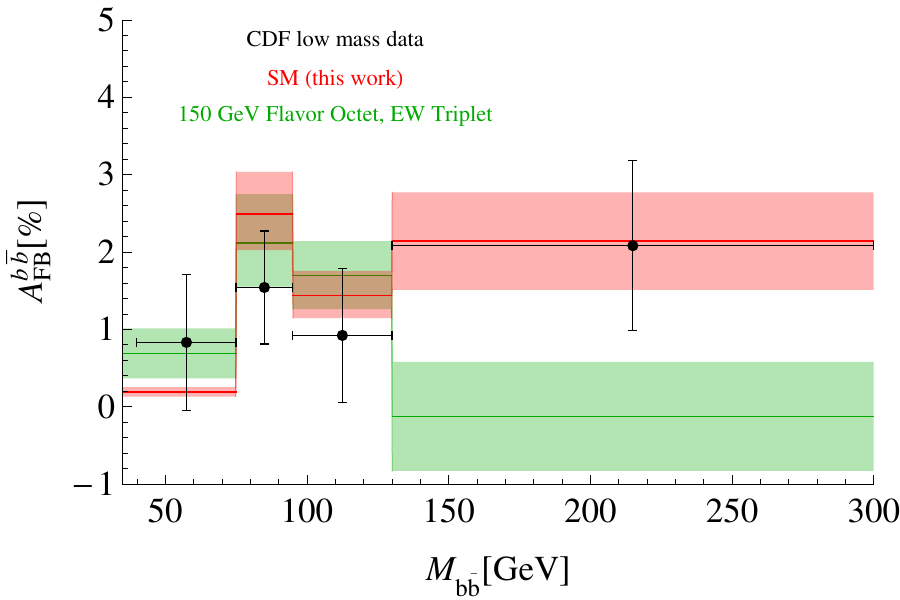}} 
   \caption{CDF's measurements of $A_{FB}^{b\bar{b}}$~\cite{CDF:2014b} are plotted in black, and the SM predictions from this work are shown in red. Plotted in brown and green respectively are predictions for the bottom asymmetry due to a 105 GeV scalar weak doublet and a flavor octet of 150 GeV EW triplet vectors.}
 \label{fig:cdflow4}
   \end{figure*}

\begin{figure*}
  \centering
 \subfloat{\label{fig:lepout}\includegraphics[width=0.5\textwidth]{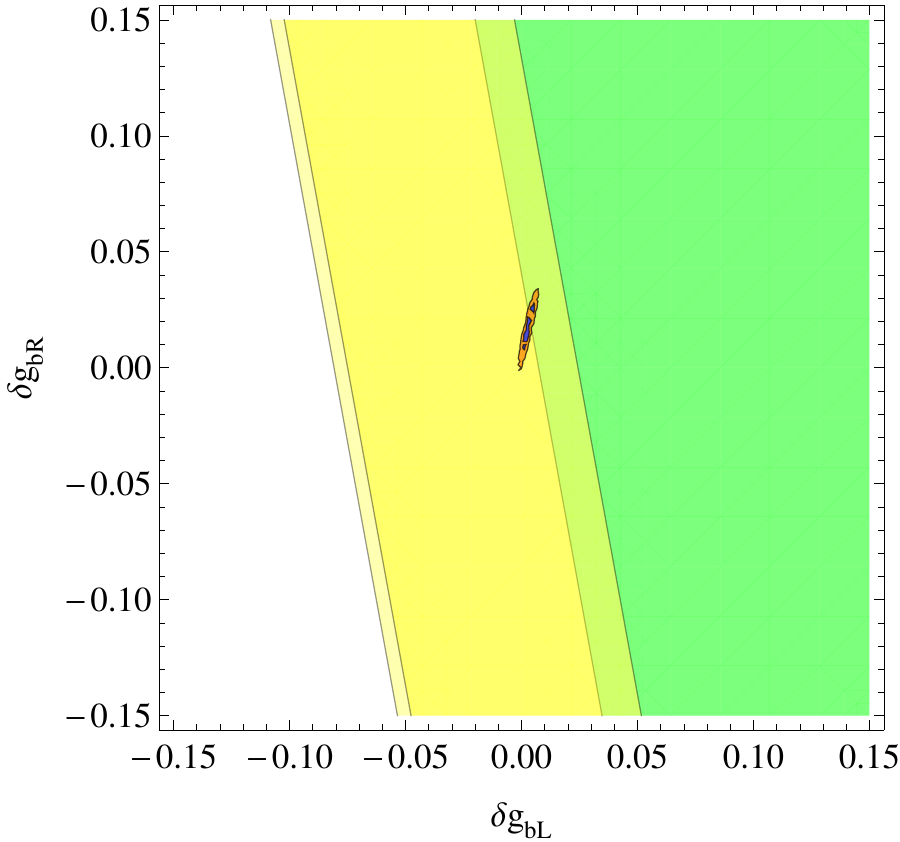}} 
  \subfloat{\label{fig:lepzoom}\includegraphics[width=0.47\textwidth]{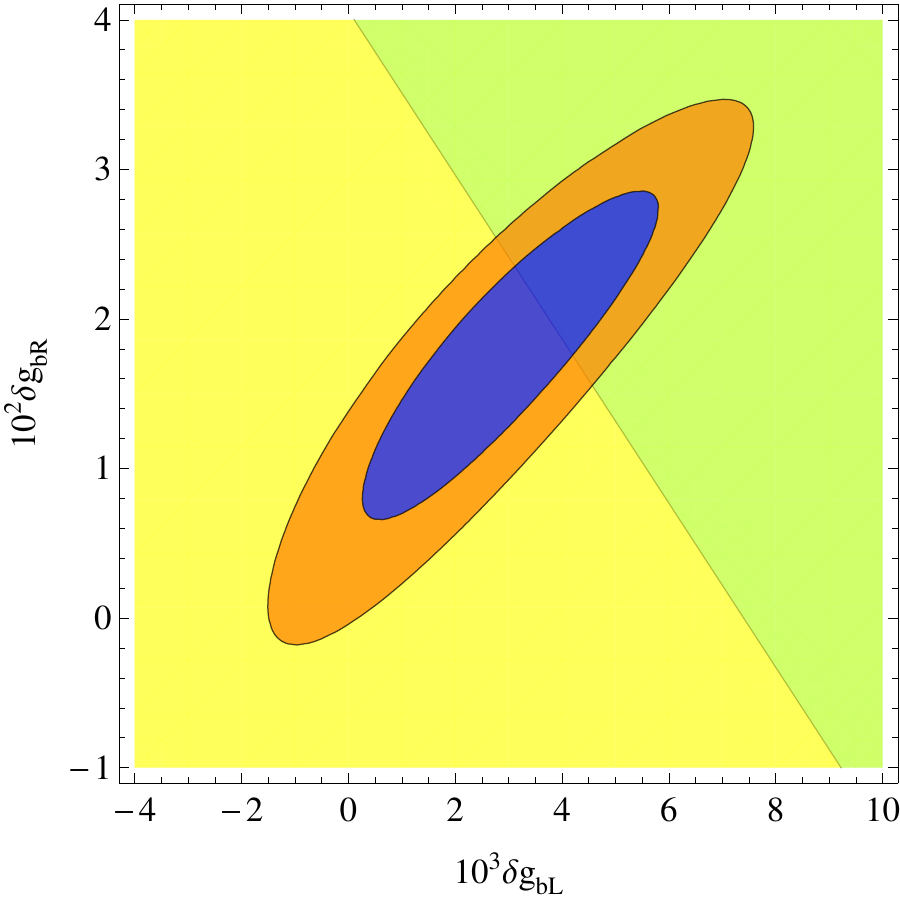}} 
   \caption{(left) Results of a fit to measurements of $A_{FB}^{(0,b)}$ and $R_b$ at LEP1.  Regions favored at the 1 and 2$\sigma$ levels are given in blue and orange respectively. Regions allowed by CDF and LHCb measurements of $A_{FB}^{b\bar{b}}$ and $A_{C}^{b\bar{b}}$ are shown in green and yellow. See the text for details on the parameter space allowed by CDF and LHCb. (right) Zoomed in version of the plot of the left, centered on the region allowed by LEP1.}
 \label{fig:lep}
   \end{figure*}

The $b\bar{b}$ forward-backward asymmetry at LEP1, $A_{FB}^{(0,b)}$, was measured to be 2.3$\sigma$ below the SM prediction~\cite{Z-Pole}. This deviation can be explained by modifying the $Zb\bar{b}$ couplings as follows,
\begin{widetext}
\begin{equation}
\mathcal{L} \supset \frac{e}{s_W c_W} Z_{\mu} \bar{b} \gamma^{\mu} \left(\left(T^3_b - Q_b s_W^2 + \delta g_{bL}\right) P_L + \left(- Q_b s_W^2 + \delta g_{bR}\right) P_R\right) b,
\end{equation}
\end{widetext}
where deviations from the SM are parameterized by $\delta g_{bL,R}$. A constraint on these modifications comes from the ratio of the partial width $Z \to b\bar{b}$ to the inclusive hadronic width of the $Z$ at LEP1, $R_b$, which is consistent with the Standard Model prediction~\cite{Z-Pole}. A two parameter fit of $\delta g_{bL,R}$ to $A_{FB}^{(0,b)}$ and $R_b$ is made using the theoretical and experimental values in~\cite{Agashe:2014kda}. Similar fits have been performed in the past, see e.g.~\cite{Batell:2012ca}. The regions favored by the fit at the 1$\sigma$ and 2$\sigma$ levels are given in Fig.~\ref{fig:lep} in blue and orange respectively. CDF and LHCb have made measurements of $A_{FB}^{b\bar{b}}$ and $A_{C}^{b\bar{b}}$ near the $Z$-pole, which also constrains the parameters $\delta g_{bL,R}$. In Fig.~\ref{fig:lepout} (left panel), the darker green and darker yellow regions correspond to values of $\delta g_{bL,R}$ that are consistent with both the CDF and the LHCb measurements at the 1$\sigma$ and 2$\sigma$ levels respectively. The lighter green region is allowed at 1$\sigma$ by CDF and 2$\sigma$ by LHCb. Lastly, the lighter yellow region is allowed by CDF at 2$\sigma$.  Fig.~\ref{fig:lepzoom} shows a zoomed in version of~\ref{fig:lepout}, centered on the region allowed by LEP1.

Fig.~\ref{fig:lep} shows that the $Zb\bar{b}$ couplings can be modified to explain the anomalously low $A_{FB}^{(0,b)}$ while being consistent with the $b\bar{b}$ asymmetry measurements at hadron colliders. This result is perhaps not so surprising since the measurements by CDF and LHCb have fairly large uncertainties. However, given the far larger amount of data expected during Run-2 of the LHC, it may possible for LHCb to constrain the parameter space for possible explanations of $A_{FB}^{(0,b)}$. The CDF measurement favors slightly smaller values of $\delta g_{bL}$ for a given value of $\delta g_{bR}$ than the LHCb measurement does. However, the width of the bands allowed by CDF and LHCb are about the same size. 

\section{Conclusions}
\label{sec:con}
The preliminary results of CDF at both high and low invariant mass are consistent with the SM predictions made in this work and in Ref.~\cite{Grinstein:2013mia}. The predictions of~\cite{Grinstein:2013mia} were expanded on in this work to include the mixed EW-QCD corrections in an approximate way, which were found to be small in magnitude. 

The charge asymmetry at 7 TeV measured by LHCb is found to be in good agreement with Standard Model. It is also predicted that the charge asymmetry at 13 \& 14 TeV will be smaller than at 7 \& 8 TeV. In addition, it was shown that $t$-channel $W$ exchange makes a non-negligible contribution to the charm-quark charge asymmetry.

Both the preliminary results of CDF at low mass and LHCb results at 7 TeV include a measured asymmetry in the bin containing the $Z$-pole that is larger than the asymmetry in the adjacent invariant mass bins, as predicted in this work and~\cite{Grinstein:2013mia}. 

D0's result for $A_{FB}$ is consistent with zero, and with the prediction of a very small asymmetry made in this work. On the other hand, the prediction for the inclusive asymmetry made by D0 using MC@NLO+Herwig is 3.3 standard deviations above what was observed.

Several BSM scenarios proposed for $A_{FB}^{t\bar{t}}$, including an 100 GeV axigluon, are ruled out by this combination of SM predictions and measurements. On the other hand, it was shown that the $Zb\bar{b}$ couplings can be modified to explain the 2.3$\sigma$ anomaly in $A_{FB}^{(0,b)}$ at LEP1 while being consistent with the $b\bar{b}$ asymmetry measurements at hadron colliders.

\begin{acknowledgments}
We thank Ben Grinstein for past collaboration, many helpful conversations, and comments on the manuscript. We would also like to thank Pavol Bartos, Julie Hogan, Michael Williams, and Jon Wilson for helpful discussions regarding experimental results. Additionally, we have benefited from conversations with Rhorry Gauld, Ulrich Haisch, Ken Intriligator, Aneesh Manohar, and Alexey Matsedonskyi. This work was supported in part by MIUR-FIRB grant RBFR12H1MW.
\end{acknowledgments}


%
%
\end{document}